# Thermal Transport of GaN/Substrate Heterostructures under Non-Uniform Heat Source

Ershuai Yin[*], Wenzhu Luo, Lei Wang, Enjian Sun, Qiang Li[*]

[a] MIIT Key Laboratory of Thermal Control of Electronic Equipment, School of Energy and Power Engineering,

Nanjing University of Science & Technology, Nanjing, Jiangsu 210094, China

**Abstract:** Heat generated in gallium nitride (GaN) high-electron-mobility transistors (HEMTs) is often concentrated in nanoscale regions and must dissipate through multiple heterostructures. However, the influence of non-uniform heat sources on the thermal transport of such heterostructures remains unclear. In this work, a thermal transport model for heterostructures under the non-uniform heat source is developed by combining first-principles calculations with Monte Carlo simulations. Temperature, heat flux, and spectral thermal conductance distributions are compared between uniform and non-uniform heat sources. The effects of heterostructure height, heat source width, and heat source height on thermal transfer characteristics are analyzed for four typical heterostructures: GaN/AlN, GaN/Diamond, GaN/Si, and GaN/SiC. The results reveal that non-uniform heat sources have little effect on average interfacial thermal conductance but induce pronounced local non-uniformity when the heterostructure height is small. The interfacial thermal conductance near the heat source region is significantly higher than that in other areas. As the heat source non-uniformity increases, the total thermal resistance of the heterostructure rises markedly, reaching several times that under uniform heat sources. Finite-element calculations fail to capture the combined effects of non-uniform heating and microscale dimensions, leading to a severe underestimation of heterostructure total thermal resistance. This work reveals the thermal transport mechanisms of heterostructures under non-uniform heat sources and provides theoretical guidance for the thermal design of wide-bandgap semiconductor devices.

---

[*] Corresponding authors. E-mail address: yes@njust.edu.cn (E. Yin), liqiang@njust.edu.cn (Q. Li)

# 1 Introduction

Wide-bandgap semiconductors, represented by gallium nitride (GaN), possess large breakdown voltage, high thermal stability, and high operating frequency, making them ideal materials for high-voltage and high-frequency electronic devices [1]. However, as chips move toward higher power densities and smaller dimensions, self-heating effects lead to severe local heat accumulation and hot-spot formation, creating critical bottlenecks for device performance and reliability [2]. Efficiently removing the heat generated inside chips and reducing local hot-spot temperatures have therefore become central challenges in current chip thermal management [3,4].

One major factor impeding heat removal from chips is interfacial thermal resistance (ITR) [5]. It mainly arises from lattice mismatch between dissimilar materials, acoustic property differences, and defects, and significantly hinders heat transfer across interfaces [6]. Experimental results showed that, in GaN transistors with diamond substrates, ITR could account for more than 40% of the total temperature rise [7]. Theoretical studies also indicated that, in GaN-based transistors, ITR may contribute over 50% of the temperature rise [8]. Therefore, elucidating the thermal transport mechanisms of heterostructures and developing effective enhancement strategies are of great importance for overcoming heat dissipation bottlenecks in high-power chips [9].

Some research has been carried out on the mechanisms and regulation of thermal transport across heterogeneous interfaces [2]. For example, Cheng et al. [10] combined Raman spectroscopy with scanning transmission electron microscopy electron energy-loss spectroscopy (STEM-EELS) to probe the phonon vibration modes at Si/Ge interfaces. They found a localized phonon mode near 12 THz within 1.2 nm of the interface, revealing additional phonon transport channels. Li et al. [11] employed the same technique to AlN/Si interfaces and identified multiple interfacial phonon modes. Extended and localized modes served as phonon bridges, linking the phonon modes of bulk AlN and Si, and significantly enhanced interfacial thermal conductance (ITC). Wu et al. [12] used molecular dynamics simulations with machine-learning interatomic potentials to study GaN/BAs interfaces and found that highly matched lattice vibrations yield an ITC of up to 260 MWm$^{-2}$K$^{-1}$. We [13] have also employed machine-learning molecular dynamics to investigate the effect of Al composition on thermal transport across GaN/Al$_x$Ga$_{1-x}$N and AlN/Al$_x$Ga$_{1-x}$N interfaces. The results showed that elemental doping promoted phonon energy redistribution and improved phonon spectrum matching, thereby enhancing interfacial heat transfer. Compared with the ITC of the GaN/AlN interface, the ITCs of GaN/Al$_{0.5}$Ga$_{0.5}$N and AlN/Al$_{0.5}$Ga$_{0.5}$N increased by 128% and 229%, respectively. In recent years, several approaches such as phonon-bridge interlayers [14,15], nanostructured interfaces [16,17], and defect engineering [18,19] have also been proposed, providing alternative strategies for interfacial thermal transport enhancement.

Another reason for thermal accumulation in chips is extremely non-uniform heat generation [20].

In a typical GaN HEMT, heat is mainly generated in the channel layer beneath the gate. Most of the heat is concentrated within a narrow region about 100 nm wide and 1 nm thick [21]. Compared with the channel thickness of 1~3 μm and widths of tens of micrometers, the heat source region is extremely small [22]. Such small heat source dimensions make it difficult for traditional Fourier thermal conduction models and ballistic transport models with gray-medium approximations to accurately describe the complex phonon scattering behavior near nanoscale heat sources and multilayer structures. Significant local temperature gradients and enhanced phonon scattering may alter phonon energy distributions, leading to locally strengthened or weakened ITC. However, existing studies on interfacial thermal transfer generally assume uniform heat sources or employ isothermal boundaries, with little consideration of non-uniform heating conditions. In addition, when heat spreads from a confined heat source region into the much larger substrate, significant spreading resistance is introduced near the junction [23], which becomes a dominant factor limiting heat dissipation [24]. The influence of nanoscale non-uniform heat sources on thermal spreading in heterostructures, however, remains rarely explored.

Therefore, this work develops a thermal transport model for heterostructures under the non-uniform heat source by combining first-principles calculations with Monte Carlo simulations. Using the GaN/AlN heterostructure as an example, Temperature, heat flux, and spectral thermal conductance distributions under uniform and non-uniform heat sources are first compared. The effects of structure height, heat source width, and heat source height on interfacial thermal transport are analyzed. Then, the influence of heat source non-uniformity on GaN/AlN, GaN/Diamond, GaN/Si, and GaN/SiC heterostructures is analyzed. Finally, the total thermal resistance and temperature distributions obtained from macroscopic finite-element calculations are compared with those from Monte Carlo simulations.

## 2 Methodology

Figure 1 shows the schematic diagram of the GaN/substrate heterostructure under the non-uniform and uniform heat sources. The substrate materials studied include diamond, SiC, Si, and AlN. The total heterostructure height is $H$, with width $W$ and depth $D$. The GaN and substrate layers each occupy half of the total height, so their individual heights are $H/2$. Both the width and depth of the heterostructure are fixed at 100 nm, while the total height varies from 10 nm to 500 nm. Periodic boundary conditions are applied to all four side surfaces in the $x$ and $z$ directions to eliminate finite-size effects on vertical heat flux. The heat source is located at the top of the GaN layer. Its width and height are denoted as $w_g$ and $h_g$, respectively, and its depth is consistent with that of the heterostructure [25]. When $w_g / W < 1$, the heat source is defined as non-uniform, whereas when $w_g / W = 1$, it is defined as uniform. The top surface is set as an adiabatic boundary, and the bottom surface is maintained at a constant temperature.

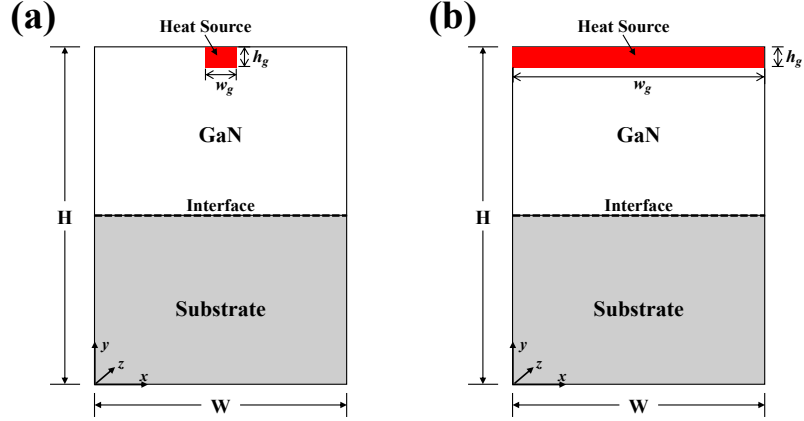

Fig.1. The schematic diagram of the GaN/substrate heterostructure under the (a) non-uniform (b) uniform heat source.

A thermal transport model for heterostructures under the non-uniform heat source is developed by combining first-principles calculations with Monte Carlo (MC) simulations. The core idea is to obtain full-band phonon properties of five materials from the first-principles calculations, and then use the MC method to solve the Boltzmann transport equation (BTE) to simulate phonon transport inside the heterostructure. We employ a variance-reduced MC method [26] to solve the deviational energy-based BTE under the relaxation time approximation (RTA) [27]. A detailed description of this approach can be found in our previous studies [14,17]. For phonon-interface scattering, the phonon spectral transmittance is defined by the diffuse mismatch model (DMM) [28]:

$$\tau_{1\to 2}(\omega') = \frac{\Delta V_2 \sum_{\mathbf{k},p} |\mathbf{v}_{g,2} \cdot \mathbf{n}| \delta(\omega' - \omega)}{\Delta V_1 \sum_{\mathbf{k},p} |\mathbf{v}_{g,1} \cdot \mathbf{n}| \delta(\omega' - \omega) + \Delta V_2 \sum_{\mathbf{k},p} |\mathbf{v}_{g,2} \cdot \mathbf{n}| \delta(\omega' - \omega)} \quad (1)$$

where subscript 1 denotes GaN and subscript 2 denotes the substrate. $\omega$ is the phonon frequency, $\Delta V$ represents the volumes of the discretized cells corresponding to the Brillouin zones, $\mathbf{v}_g$ is the phonon velocity, $\mathbf{n}$ is the unit normal vector of the interface, and $\delta$ is the Dirac delta function.

During the simulation, phonons are emitted from the volumetric heat source and the isothermal boundary. For the volumetric heat source, the effective deviational energy is defined as:

$$Q_{heat\_source} = P \times D \quad (2)$$

where $P$ is the line power density of the heat source. When phonons are emitted from the volumetric heat source, their emission positions and angles are randomly determined, and the emission probability is uniform over the entire spherical surface. The total heat source power is kept constant, maintaining a line power of 0.1 W/mm. The magnitude of the applied power does not affect phonon interfacial transport or spreading resistance but improves the accuracy of variance-reduced MC simulations [21].

All full-band phonon parameters required for the MC calculations and the DMM model, including phonon group velocity, relaxation time, and volumetric heat capacity, are obtained from first-principles calculations. Specifically, the second- and third-order force constants of the studied materials are computed using the Vienna Ab initio Simulation Package (VASP) [29,30], PHONOPY [31], and the thirdorder.py package [32]. These force constants are then used in the almaBTE package [33] to calculate all necessary phonon properties. Detailed procedures and computational settings are provided in our previous work [14]. The thermal conductivities of GaN, Si, SiC, AlN, and Diamond are calculated in the 300~500 K temperature range and compared with reported experimental data, effectively validating both the first-principles approach and the accuracy of the full-band phonon properties [14,18].

During the MC simulations, the total number of phonons is set to $6 \times 10^5$, and the equilibrium temperature is 300 K. The grid size is 100×200×20 in the $x$, $y$, and $z$ directions. The maximum number of scattering events for a single phonon is limited to $1 \times 10^5$, a value chosen through testing to balance computational efficiency and accuracy [26]. After the MC calculations are completed, the temperature and heat-flux distributions of the heterostructure are obtained. The interfacial thermal conductance is then calculated as [34,35]:

$$\text{ITC} = \frac{q}{\Delta T} \tag{3}$$

where $q$ is the heat flux across the interface, and $\Delta T$ is the interfacial temperature drop. The average ITC is calculated using the average heat flux and temperature drop across the entire interface, while the local ITC is calculated using the local heat flux and local temperature drop.

The total thermal resistance of the heterostructure is defined to evaluate its heat spreading and transfer capability [36]:

$$R = \frac{\overline{T}_s - T_c}{P} \tag{4}$$

where $\overline{T}_s$ is the average temperature of the heat source region, and $T_c$ is the temperature of the bottom cold side.

To describe the influence of heat source non-uniformity on the thermal transport characteristics of the heterostructure, the ratio of the total thermal resistance under non-uniform heating to that under uniform heating is defined as:

$$r = \frac{R_{non-uniform}}{R_{uniform}} \tag{5}$$

where $R_{non-uniform}$ is the total thermal resistance of the heterostructure under non-uniform heat sources, and $R_{uniform}$ is the total thermal resistance under uniform heating.

## 3 Results and Discussion

*3.1 Model Validation*

To verify the accuracy of the constructed model, the ITCs of GaN/Diamond, GaN/SiC, GaN/AlN, and GaN/Si interfaces are calculated under uniform heat sources and compared with experimental and non-equilibrium molecular dynamics (NEMD) results reported in the literature. The results are shown in Figure 2. For GaN/Diamond and GaN/AlN interfaces, the calculated ITCs agree very well with both experimental and NEMD values, demonstrating high accuracy. For the GaN/SiC and GaN/Si interfaces, the calculated ITCs are slightly higher than the experimental values, indicating that the fabricated heterostructures deviate from the ideal contact conditions assumed in the theoretical model. These results confirm that the model is reliable and suitable for subsequent investigations.

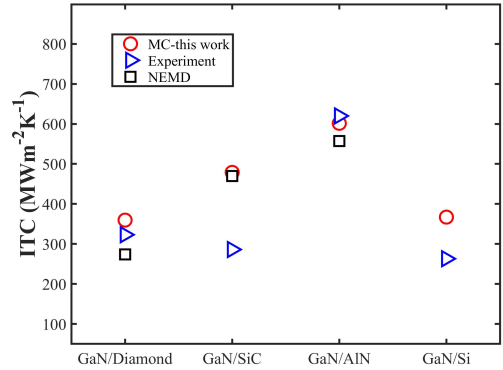

Fig. 2. Model validation. Experimental data are from Ref. [5]. NEMD results for GaN/Diamond, GaN/SiC, and GaN/AlN interfaces are from Refs. [16], [37], and [13], respectively.

*3.2 Effect of the Non-Uniform Heat Source on Heterostructure Thermal Transport*

The GaN/AlN heterostructure is used as an example to investigate the influence of non-uniform heat sources on heterostructure thermal transport. Figure 3 presents the temperature distributions for heterostructure heights H of 10, 50, 100, and 300 nm under uniform and non-uniform heat sources. The uniform heat source has a size of 100 nm × 1 nm ($w_g / W = 1$), while the non-uniform heat source is 20 nm × 1 nm ($w_g / W = 0.2$). The results show that non-uniform heat sources produce markedly different temperature distributions, and the difference is more pronounced at smaller structure heights. When the structure height is 10 nm, the high-temperature region caused by the non-uniform heat source is confined to a narrow area, with heat spreading only across a width of about 20 nm. The main reason is that, as shown in Figure 4, lateral heat spreading requires sufficient vertical space. At a height of 10 nm, most phonons are scattered at the heterointerface, leaving only about 5 nm of space for lateral heat spreading. Although a lateral heat flux as high as 6 GWm$^{-2}$ occurs, heat remains concentrated in the central region. As the structure height increases, this confinement effect rapidly weakens. From Figure 3(b), it can be observed that GaN requires a height greater than 10 nm for effective lateral heat spreading. Figure 4(c) and (d), for the case of H = 300 nm, show that when the height is sufficient, heat quickly spreads laterally near the local heat source and then transfers vertically in a uniform manner, with negligible lateral spreading in most of the region. Therefore, when the structure height is sufficiently large, non-uniform heat sources have little influence on interfacial heat transfer capability.

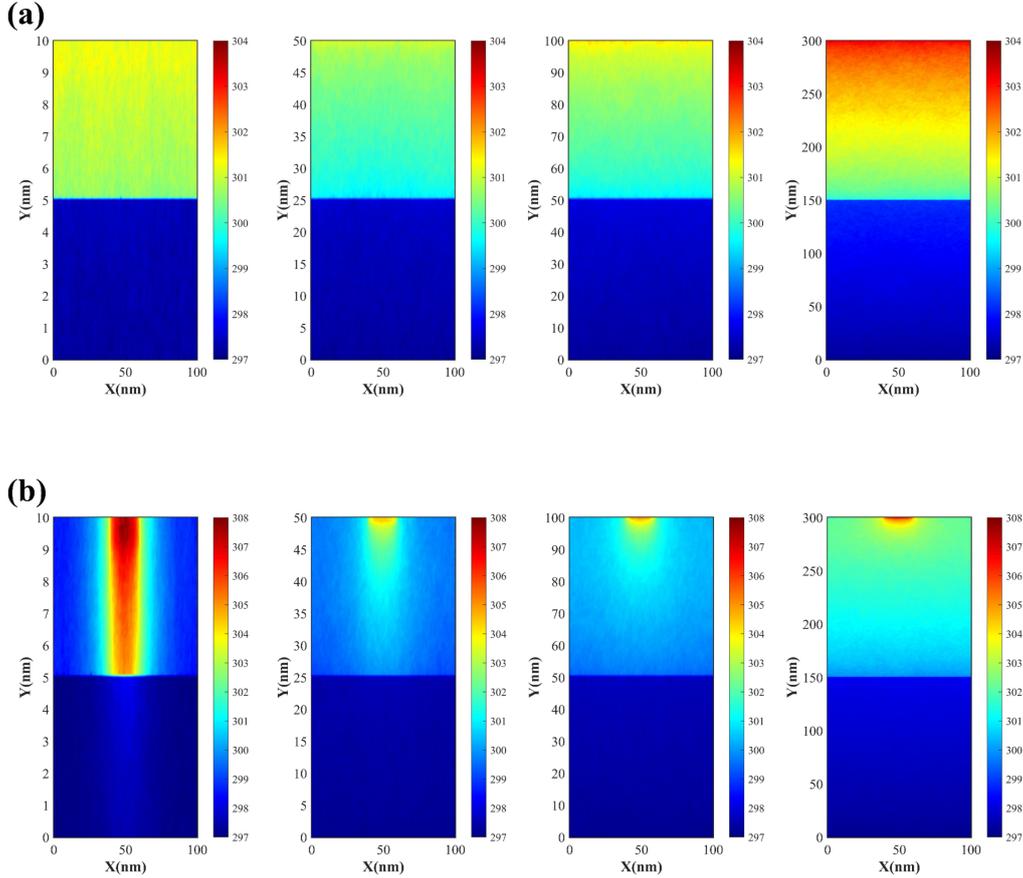

Fig. 3. Effect of GaN/AlN heterostructure height and heat source width on temperature distribution when $h_g$ = 1 nm. Each column corresponds to H = 10, 50, 100, and 300 nm. (a) Uniform heat source, $w_g / W$ = 1. (b) Non-uniform heat source, $w_g / W$ = 0.2.

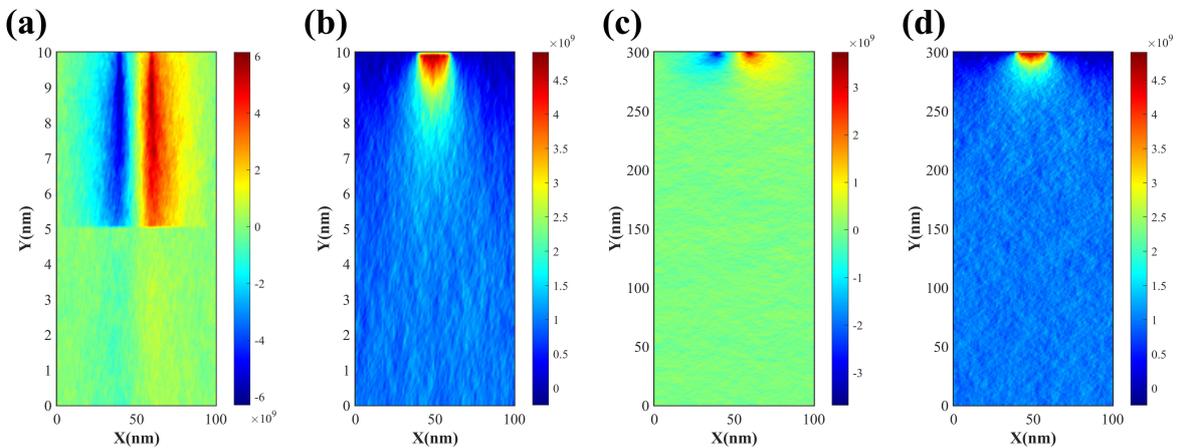

Fig. 4. Heat flux comparison in the GaN/AlN heterostructure with a heat source width of 20 nm and height of 1 nm. (a) *H* = 10 nm, heat flux in the x-direction. (b) *H* = 10 nm, heat flux in the y-direction. (c) *H* = 300 nm, heat flux in the x-direction. (d) *H* = 300 nm, heat flux in the y-direction.

Figure 5 compares the spectral thermal conductance distributions of GaN/AlN heterostructures with different heights when the heat source width is 20 nm. When the heterostructure height is small, heat transfer primarily relies on two phonon frequency ranges: 3~5 THz and 15~23 THz. According to the diffuse mismatch model, only phonons with matching frequencies on both sides of the interface can transmit. Because of the small dimensions, phonons primarily undergo ballistic transport and rarely experience three-phonon scattering to change frequency. As a result, the heat flux in the bulk is strongly influenced by interfacial phonon transmission. Figure 6 shows the phonon density of states (PDOS) of GaN and AlN and the transmittance of the GaN/AlN interface. The PDOS overlap in the 3~5 THz and 15~23 THz ranges, allowing elastic phonons to transmit. The transmittance is determined by the phonon velocities of both sides. Heat is mainly carried by acoustic phonons in the 3~5 THz range, because low-frequency acoustic phonons have higher group velocities and longer mean free paths, enabling repeated interface scattering until transmission occurs. In contrast, high-frequency optical phonons have shorter mean free paths and are more likely to undergo three-phonon scattering, which converts them to other frequencies and lowers their heat transport. As the heterostructure height increases, the spectral thermal conductance changes significantly. In GaN, the heat carried by phonons above 7.5 THz gradually decreases, while the contribution near 7.5 THz increases. In AlN, the heat carried by phonons near 9 THz and above 15 THz decreases, while the contribution of phonons in the 10~15 THz range increases. These trends occur because the influence of interfacial phonon transmission on thermal conduction inside the bulk diminishes with increasing structure height. The phonon heat transport becomes increasingly determined by the intrinsic phonon properties of the bulk materials.

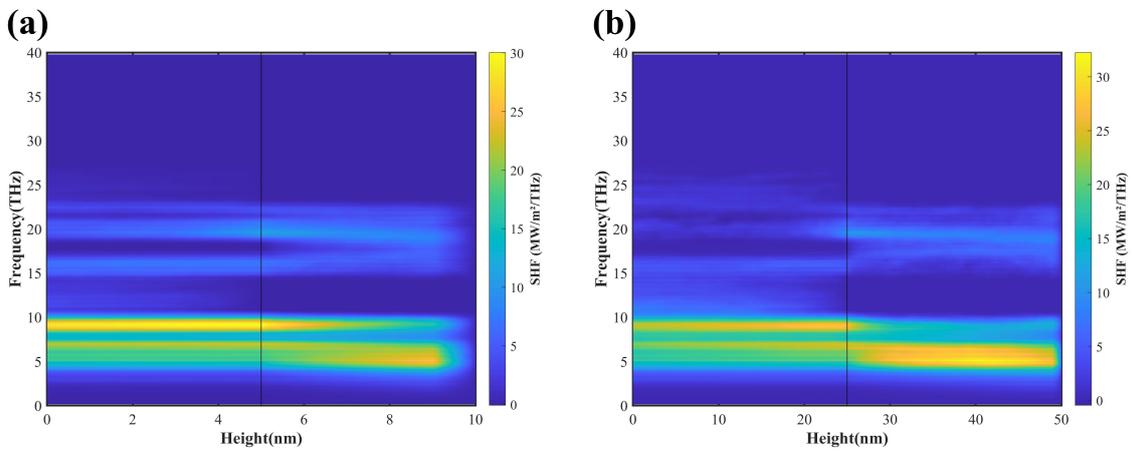

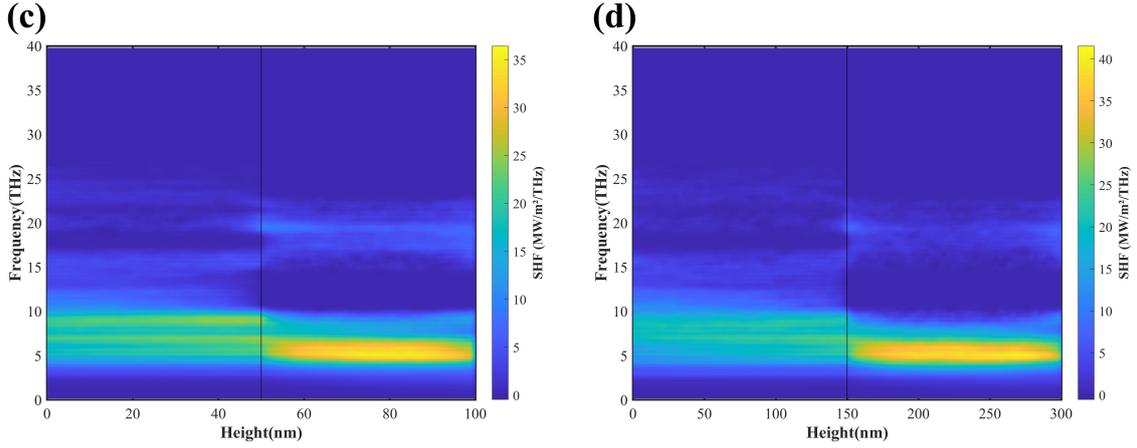

Fig. 5. Comparison of spectral thermal conductance distributions in GaN/AlN heterostructures with the heat source width of 20 nm and height of 1nm (a) $H$ = 10 nm, (b) $H$ = 50 nm, (c) $H$ = 100 nm, and (d) $H$ = 300 nm.

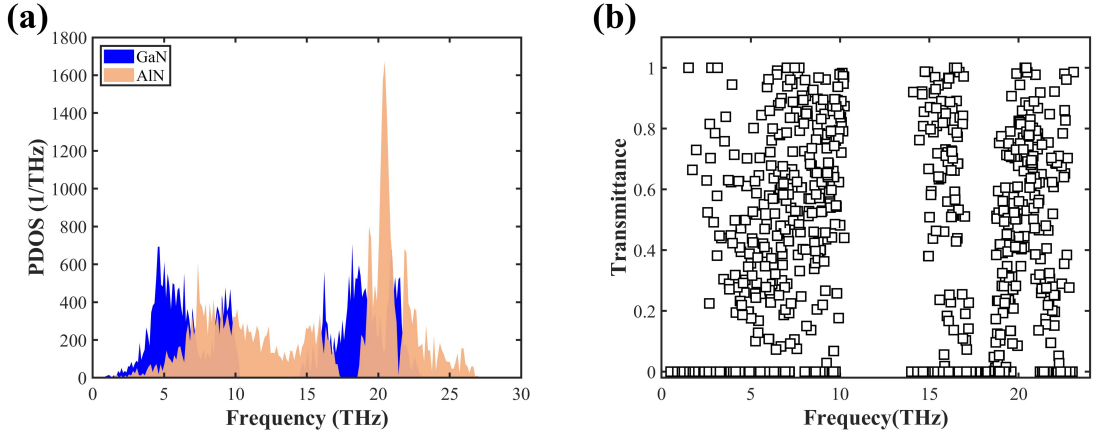

Fig. 6. (a) Phonon density of states of GaN and AlN. (b) Spectral phonon transmittance of the GaN/AlN interface.

Figure 7(a) shows the effects of structure height and heat source non-uniformity on the total thermal resistance of the GaN/AlN heterostructure when the heat source height is 10 nm. For all heat source conditions, the total thermal resistance decreases first and then increases as the structure height grows, with the increase rate gradually slowing. This behavior arises because, at small structure heights, ballistic phonon transport dominates, and phonons cannot undergo sufficient three-phonon scattering. This strongly limits phonon transmission across the interface and markedly increases interfacial thermal resistance. This phenomenon can be observed in the spectral thermal conductance contributions in Figure 5(a). Figure 7(b) presents the average interfacial thermal conductance, defined as the ratio of transmitted heat flux to the temperature difference between the upper and lower surfaces of the interface. At small structure heights, interfacial heat transfer is much weaker than at larger heights, and the interface dominates the overall heat spreading process, resulting in high total thermal resistance. As structure height continues to increase, the ITC rises rapidly and then saturates. The interface no longer limits phonon transfer, but the larger bulk height increases intrinsic thermal resistance, causing the total thermal resistance to continue rising. Comparing the average ITC values under different heat source inhomogeneities reveals that the average ITC values are largely unaffected by the heat source distribution. However, Figures 7(c) and 7(d) demonstrate that when the heat source is non-uniform and the structure height is small, the local ITC exhibits strong non-uniformity. The local ITC near the center and close to the heat source is significantly higher than that in surrounding regions, while the ITC in other areas is lower. As the structure height increases or the heat source becomes more uniform, this non-uniform distribution of local ITC is strongly weakened, and the interfacial thermal transfer becomes more uniform. Therefore, non-uniform heat sources mainly affect the spatial distribution of local ITC, while their influence on the average interfacial heat transfer property is minimal and occurs only when the heat source is highly non-uniform and the heterostructure height is small.

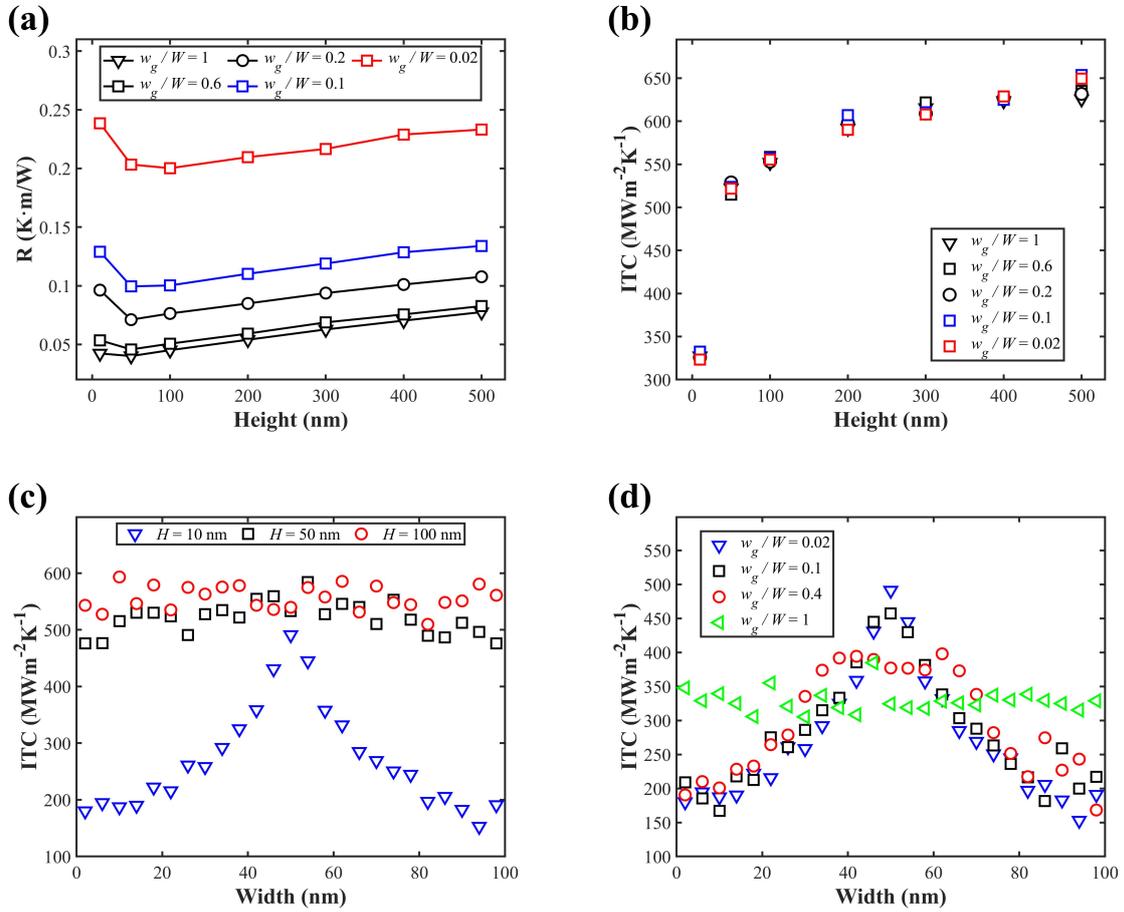

Fig. 7. Effects of heterostructure height and heat source non-uniformity on the (a) total thermal resistance, (b) average ITC. (c) Local ITC at different heterostructure heights with $w_g / W = 0.02$. (d) Effect of heat source non-uniformity on local ITC with $H = 10$ nm.

Figure 8 illustrates the effect of heat source geometry on the total thermal resistance of GaN/AlN heterostructures, with the heat source width fixed at 10 nm. As shown in Figure 8(a), the total thermal resistance decreases as the height-to-width ratio ($h_g / w_g$) increases. This trend is consistent for structures of different heights: the total thermal resistance first decreases rapidly and then more slowly as $h_g / w_g$ increases. When $H$ is 50 nm, increasing $h_g / w_g$ from 0.1 to 1 reduces the total thermal resistance by 38%. Figures 8(b) and 8(c) show the temperature distributions of GaN/AlN heterostructure with $H$ of 50 nm at $h_g / w_g$ = 0.1 and 1, respectively. Compared with the small $h_g / w_g$, increasing the heat source height mainly lowers the temperature in the heat source region. A small heat source tends to create localized high-temperature hot spots. Therefore, the primary factor affecting the total thermal resistance of heterostructures is the size of the heat source, and the larger heat sources generally result in a lower total thermal resistance.

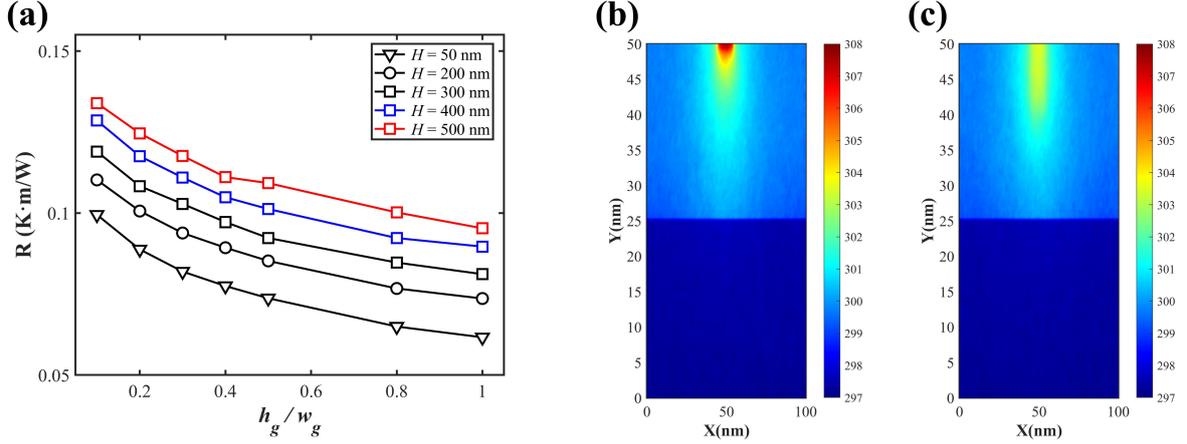

Fig.8. (a) Effects of heat source geometry and heterostructure height on total thermal resistance. Temperature distribution of the GaN/AlN heterostructure when $h_g / w_g$ is (b) 0.1, (c) 1.

At a heterostructure height of 200 nm, the influence of the heat source non-uniformity on the total thermal resistance of GaN heterostructures with different substrate materials is calculated, as shown in Figure 9. The results show that the effect of heat source non-uniformity is similar for all substrates. Specifically, as $w_g / W$ decreases, $r$ increases sharply; as $h_g$ decreases, $r$ increases further, and the influence of $w_g / W$ on $r$ becomes stronger. Comparing $r$ among different substrates reveals that non-uniform heat sources have a larger impact on heterostructures with AlN and SiC substrates. This is because GaN/AlN and GaN/SiC interfaces exhibit higher ITC than GaN/Diamond and GaN/Si interfaces. Larger contact thermal resistance leads to a higher total thermal resistance and a larger share of contact resistance. Since a high structure height reduces the effect of non-uniform heat sources on interfacial ITC, the influence of non-uniform heating on GaN/Diamond and GaN/Si heterostructures is smaller. However, due to their higher thermal contact resistance, the total thermal resistances of GaN/Diamond and GaN/Si remain larger than those of GaN/AlN and GaN/SiC. For example, when the heat source size is 10 nm × 10 nm, the total thermal resistances of GaN/Diamond and GaN/Si heterostructures are 0.085 KmW$^{-1}$ and 0.093 KmW$^{-1}$, respectively, whereas those of GaN/AlN and GaN/SiC heterostructures are only 0.073 KmW$^{-1}$ and 0.077 KmW$^{-1}$.

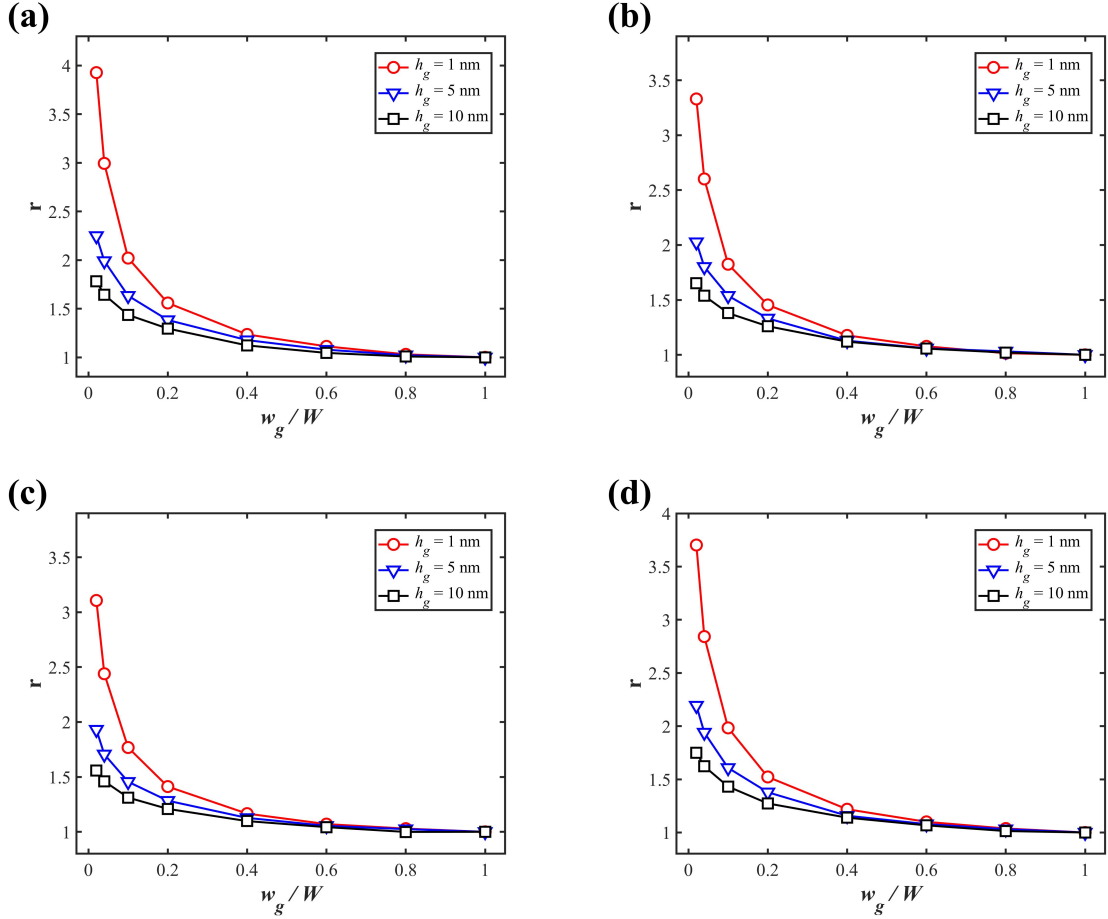

Fig. 9. Effect of the heat source non-uniformity on GaN heterostructures with different substrates at a fixed structure height of 200 nm. (a) AlN substrate. (b) Diamond substrate. (c) Si substrate. (d) SiC substrate.

*3.3 Comparison of Results with the FEM Method*

Figure 10 compares the macroscopic FEM results with those of the MC method. Figure 10(a) shows that FEM underestimates the total thermal resistance of the heterostructure for all structure heights. The main reason is that FEM assumes purely diffusive heat transport and cannot capture ballistic phonon transport at reduced dimensions. This leads to an overestimation of heat transfer inside the bulk material and near the interface. As a result, even when the heat source is uniform ($w_g / W = 1$), FEM predicts a total thermal resistance significantly lower than that calculated by the MC method. Another important observation is that FEM fails to represent the effect of heat source size on total thermal resistance. When $w_g / W$ decreases from 1 to 0.02, FEM shows little change in total resistance. In contrast, MC results indicate that decreasing $w_g / W$ and increasing heat source non-uniformity cause a rapid rise in total resistance, eventually leading to several-fold increases. Figures 10(b) and 10(c) compare the temperature distributions predicted by FEM and MC for $H = 100$ nm, $h_g = 1$ nm, and $w_g / W = 1$. The FEM calculation shows heat spreading evenly outward, yielding a relatively uniform temperature field and a lower heat source temperature. In contrast, the MC calculation reveals a distinct hot spot, with an average heat source temperature of 307.0 K compared to only 299.5 K predicted by FEM. These results demonstrate that traditional macroscopic FEM significantly underestimates the total thermal resistance and is therefore not applicable when nanoscale heat sources or structures are present.

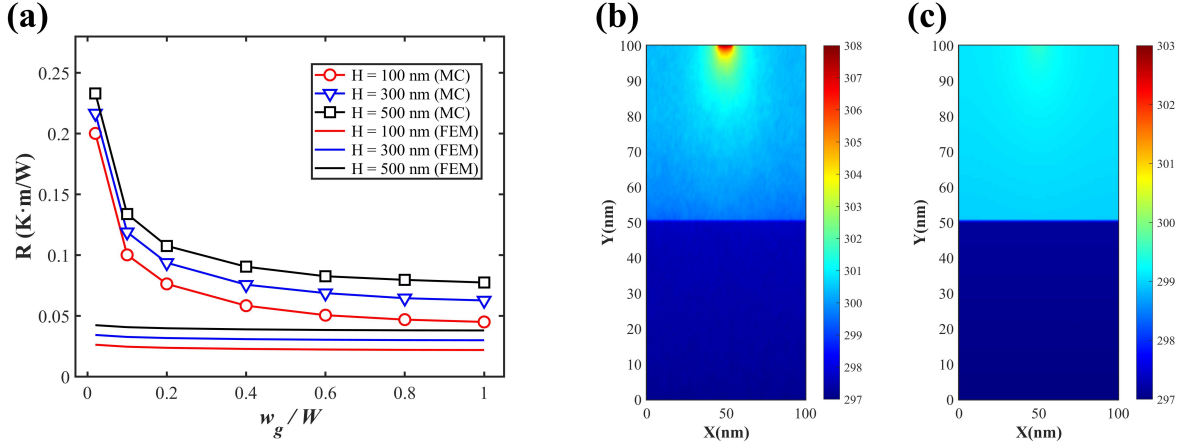

Fig. 10. Comparison of MC and FEM results for the GaN/AlN interface at $h_g = 1$ nm and $H = 100$ nm. (a) Total thermal resistance under different $w_g / W$. (b) Temperature distribution calculated by the MC method at $w_g / W =$ 0.1. (c) Temperature distribution calculated by the FEM method at $w_g / W = 0.1$.

## 4 Conclusions

This work develops a thermal transport model for heterostructures under the non-uniform heat source by combining first-principles calculations with Monte Carlo simulations. The effect of non-uniform heat sources on the thermal transport of GaN/substrate heterostructures is investigated. The main conclusions are as follows:

1) Non-uniform heat sources have little effect on the average interfacial thermal conductance but create significant local interfacial thermal conductance non-uniformity when the structure height is small. The interfacial thermal conductance near the heat source region is much higher than that in other regions.

2) The total thermal resistance of the heterostructure decreases first and then increases with increasing structure height. At small heights, ballistic phonon transport dominates, heat transfer relies mainly on low-frequency acoustic phonons, the interfacial thermal conductance is low, and the total resistance is high. As the height increases, the influence of interfacial thermal resistance weakens, while the intrinsic bulk resistance becomes dominant, gradually increasing the total thermal resistance.

3) As the heat source non-uniformity increases, the total thermal resistance rises significantly, reaching several times the value under uniform heat sources.

4) Conventional finite element method calculations cannot capture the effects of non-uniform heat sources and nanoscale dimensions, leading to severe underestimation of the total thermal resistance of heterostructures.

This work reveals the thermal transport mechanisms of heterostructures under non-uniform heat sources and provides theoretical guidance for the thermal design of wide-bandgap semiconductor devices.

**Acknowledgments**

This work was supported by the National Natural Science Foundation of China (Grant NO. 92473204) and the Fundamental Research Funds for the Central Universities (No.30923010917).

**References**

[1] Cheng Z, Huang Z, Sun J, Wang J, Feng T, Ohnishi K, et al. (Ultra)wide bandgap semiconductor heterostructures for electronics cooling. Appl Phys Rev 2024;11:041324. https://doi.org/10.1063/5.0185305.

[2] Zhang G, Dong S, Xin Q, Guo L, Wang X, Xin G, et al. Transistor-level thermal management in wide and ultra-wide bandgap power semiconductor transistors: A review. Int J Therm Sci 2026;219:110200. https://doi.org/10.1016/j.ijthermalsci.2025.110200.

[3] Tang D-S, Cao B-Y. Phonon thermal transport and its tunability in GaN for near-junction thermal management of electronics: A review. Int J Heat Mass Transf 2023;200:123497. https://doi.org/10.1016/j.ijheatmasstransfer.2022.123497.

[4] Zhou J, Zhong L, Feng X, Zhang W, Liu X, Zhou H, et al. Recent Advances in Device-Level Thermal Management Technologies for Wide Bandgap Semiconductor: A Review. IEEE Trans Electron Devices 2025;72:2769–82. https://doi.org/10.1109/TED.2025.3562506.

[5] Feng T, Zhou H, Cheng Z, Larkin LS, Neupane MR. A Critical Review of Thermal Boundary Conductance across Wide and Ultrawide Bandgap Semiconductor Interfaces. ACS Appl Mater Interfaces 2023;15:29655–73. https://doi.org/10.1021/acsami.3c02507.

[6] Zhao X, Hu W. Progress in the semiconductor/diamond heterogeneous integrations: Technical methods, interfacial phonon transport, and thermal characterizations. Surf Interfaces 2024;46:104178. https://doi.org/10.1016/j.surfin.2024.104178.

[7] Pomeroy JW, Bernardoni M, Dumka DC, Fanning DM, Kuball M. Low thermal resistance GaN-on-diamond transistors characterized by three-dimensional Raman thermography mapping. Appl Phys Lett 2014;104:083513. https://doi.org/10.1063/1.4865583.

[8] Filippov KA, Balandin AA. The Effect of the Thermal Boundary Resistance on Self-Heating of AlGaN/GaN HFETs. Mater Res Soc Internet J Nitride Semicond Res 2003;8:e4. https://doi.org/10.1557/S1092578300000478.

[9] Zhang X-D, Yang G, Cao B-Y. Bonding-Enhanced Interfacial Thermal Transport: Mechanisms, Materials, and Applications. Adv Mater Interfaces 2022;9:2200078. https://doi.org/10.1002/admi.202200078.

[10] Cheng Z, Li R, Yan X, Jernigan G, Shi J, Liao ME, et al. Experimental observation of localized interfacial phonon modes. Nat Commun 2021;12:6901. https://doi.org/10.1038/s41467-021-27250-3.

[11] Li Y, Qi R, Shi R, Hu J, Liu Z, Sun Y, et al. Atomic-scale probing of heterointerface phonon bridges in nitride semiconductor. Proc Natl Acad Sci 2022;119:e2117027119. https://doi.org/10.1073/pnas.2117027119.

[12] Wu J, Zhou E, Huang A, Zhang H, Hu M, Qin G. Deep-potential enabled multiscale simulation of gallium nitride devices on boron arsenide cooling substrates. Nat Commun 2024;15:2540. https://doi.org/10.1038/s41467-024-46806-7.

[13] Luo W, Yin E, Wang L, Lian W, Wang N, Li Q. Heat transfer enhancement of N-Ga-Al semiconductors heterogeneous interfaces. Int J Heat Mass Transf 2025;244:126902. https://doi.org/10.1016/j.ijheatmasstransfer.2025.126902.

[14] Yin E, Li Q, Luo W, Wang L. Enhancing interfacial thermal conductance in Si/Diamond heterostructures by phonon bridge 2025. https://doi.org/10.48550/arXiv.2507.22490.

[15] Yang L, Wan X, Ma D, Jiang Y, Yang N. Maximization and minimization of interfacial thermal conductance by modulating the mass distribution of the interlayer. Phys Rev B 2021;103:155305. https://doi.org/10.1103/PhysRevB.103.155305.

[16] Yin E, Li Q, Lian W. Mechanisms for enhancing interfacial phonon thermal transport by large-size nanostructures. Phys Chem Chem Phys 2023;25:3629–38. https://doi.org/10.1039/D2CP02887E.

[17] Luo W, Wang N, Lian W, Yin E, Li Q. Enhancing interfacial thermal transport by nanostructures: Monte Carlo simulations with ab initio phonon properties. J Appl Phys 2025;137. https://doi.org/10.1063/5.0243745.

[18] Yin E, Luo W, Wang L, Li Q. Effects of Defects on Thermal Transport across Solid/Solid Heterogeneous Interfaces 2025. https://doi.org/10.48550/arXiv.2508.12744.

[19] Yang C, Wang J, Ma D, Li Z, He Z, Liu L, et al. Phonon transport across GaN-diamond interface: The nontrivial role of pre-interface vacancy-phonon scattering. Int J Heat Mass Transf 2023;214:124433. https://doi.org/10.1016/j.ijheatmasstransfer.2023.124433.

[20] Chen X, Boumaiza S, Wei L. Self-Heating and Equivalent Channel Temperature in Short Gate Length GaN HEMTs. IEEE Trans Electron Devices 2019;66:3748–55. https://doi.org/10.1109/TED.2019.2926742.

[21] Li H-L, Shen Y, Hua Y-C, Sobolev SL, Cao B-Y. Hybrid Monte Carlo-Diffusion Studies of Modeling Self-Heating in Ballistic-Diffusive Regime for Gallium Nitride HEMTs. J Electron Packag 2022;145. https://doi.org/10.1115/1.4054698.

[22] Hao Q, Zhao H, Xiao Y, Kronenfeld MB. Electrothermal studies of GaN-based high electron mobility transistors with improved thermal designs. Int J Heat Mass Transf 2018;116:496–506. https://doi.org/10.1016/j.ijheatmasstransfer.2017.09.048.

[23] Muzychka YS, Culham JR, Yovanovich MM. Thermal Spreading Resistance of Eccentric Heat Sources on Rectangular Flux Channels. J Electron Packag 2003;125:178–85. https://doi.org/10.1115/1.1568125.

[24] Hua Y-C, Li H-L, Cao B-Y. Thermal Spreading Resistance in Ballistic-Diffusive Regime for GaN HEMTs. IEEE Trans Electron Devices 2019;66:3296–301. https://doi.org/10.1109/TED.2019.2922221.

[25] Hao Q, Zhao H, Xiao Y. A hybrid simulation technique for electrothermal studies of two-dimensional GaN-on-SiC high electron mobility transistors. J Appl Phys 2017;121:204501. https://doi.org/10.1063/1.4983761.

[26] Pathak A, Pawnday A, Roy AP, Aref AJ, Dargush GF, Bansal D. MCBTE: A variance-reduced Monte Carlo solution of the linearized Boltzmann transport equation for phonons. Comput Phys Commun 2021;265:108003. https://doi.org/10.1016/j.cpc.2021.108003.

[27] Zhao X, Qian X, Li X, Yang R. Thermal conductance of nanostructured interfaces from Monte Carlo simulations with ab initio-based phonon properties. J Appl Phys 2021;129:215105. https://doi.org/10.1063/5.0050175.

[28] Reddy P, Castelino K, Majumdar A. Diffuse mismatch model of thermal boundary conductance using exact phonon dispersion. Appl Phys Lett 2005;87:211908. https://doi.org/10.1063/1.2133890.

[29] Kresse G, Furthmüller J. Efficiency of ab-initio total energy calculations for metals and

semiconductors using a plane-wave basis set. Comput Mater Sci 1996;6:15–50. https://doi.org/10.1016/0927-0256(96)00008-0.

[30] Kresse G, Furthmüller J. Efficient iterative schemes for ab initio total-energy calculations using a plane-wave basis set. Phys Rev B 1996;54:11169–86. https://doi.org/10.1103/PhysRevB.54.11169.

[31] Chaput L, Togo A, Tanaka I, Hug G. Phonon-phonon interactions in transition metals. Phys Rev B 2011;84:094302. https://doi.org/10.1103/PhysRevB.84.094302.

[32] Li W, Lindsay L, Broido DA, Stewart DA, Mingo N. Thermal conductivity of bulk and nanowire $Mg_{2}Si_{x}Sn_{1-x}$ alloys from first principles. Phys Rev B 2012;86:174307. https://doi.org/10.1103/PhysRevB.86.174307.

[33] Carrete J, Vermeersch B, Katre A, Roekeghem AV, Wang T, Madsen G, et al. almaBTE : A solver of the space–time dependent Boltzmann transport equation for phonons in structured materials. Comput Phys Commun 2017.

[34] Xu Y, Wang G, Zhou Y. Broadly manipulating the interfacial thermal energy transport across the Si/4H-SiC interfaces via nanopatterns. Int J Heat Mass Transf 2022;187:122499. https://doi.org/10.1016/j.ijheatmasstransfer.2021.122499.

[35] Bergman T, Incropera F, DeWitt D, Lavine A. Fundamentals of Heat and Mass Transfer. 7th ed. New York: John Wiley & Sons; 2011.

[36] Shen Y, Hua Y-C, Li H-L, Sobolev SL, Cao B-Y. Spectral Thermal Spreading Resistance of Wide Bandgap Semiconductors in Ballistic-Diffusive Regime. ArXiv220103788 Phys 2022. https://doi.org/10.1109/TED.2022.3168798.

[37] Hu M, Zhang X, Poulikakos D, Grigoropoulos CP. Large "near junction" thermal resistance reduction in electronics by interface nanoengineering. Int J Heat Mass Transf 2011;54:5183–91. https://doi.org/10.1016/j.ijheatmasstransfer.2011.08.027.